\documentclass[twocolumn,aps,prb,amsmath,floatfix]{revtex4}
\usepackage{graphicx}
\begin{document}
\title{Mott transition in two-dimensional frustrated compounds}
\author{A.~Liebsch$^1$, H. Ishida$^2$, and J. Merino$^3$} 
\affiliation{$^1$Institut f\"ur Festk\"orperforschung, 
             Forschungszentrum J\"ulich, 52425 J\"ulich, Germany \\
      $^2$College of Humanities and Sciences, Nihon University,~Tokyo 156, Japan\\
\mbox{$^3$Departamento de F\'\i sica Te\'orica de la Materia Condensada,
             Universidad Aut\'onoma de Madrid, Madrid 28049, Spain}}
\begin{abstract}
The phase diagrams of isotropic and anisotropic triangular lattices with local 
Coulomb interactions are evaluated within cluster dynamical mean field theory. 
As a result of partial geometric 
frustration in the anisotropic lattice, short range correlations are shown to give 
rise to reentrant behavior which is absent in the fully frustrated isotropic limit. 
The qualitative features of the phase diagrams including the critical temperatures 
are in good agreement with experimental data for the layered organic charge transfer 
salts $\kappa$-(BEDT-TTF)$_2$Cu[N(CN)$_2$]Cl and $\kappa$-(BEDT-TTF)$_2$Cu$_2$(CN)$_3$.
\\
\mbox{\hskip1cm}  \\
DOI: \hfill PACS numbers: 71.10.Fd, 71.15.-m,71.27.+a
\end{abstract}
\maketitle

\section{Introduction}

The influence of spatial quantum fluctuations on the nature of the Mott 
transition in strongly correlated systems is currently of great interest. 
A class of materials in which these effects can be studied in detail are the 
layered charge transfer salts of the $\kappa$-(BEDT-TTF)$_2 X$ family, 
where $X$ denotes an inorganic monovalent anion such as Cu[N(CN)$_2$]Cl 
or Cu$_2$(CN)$_3$.
The electronic properties of these compounds have been shown to be highly 
sensitive functions of hydrostatic pressure.
\cite{komatsu,ito,lefebre,limelette,shimizu,kagawa,kurosaki,ohira}
As a result, the temperature versus pressure phase diagram is remarkably rich, 
exhibiting Fermi-liquid and bad-metallic behavior, superconductivity, as well 
as paramagnetic and antiferromagnetic insulating phases. 
These observations suggest fascinating connections to analogous phenomena 
in various transition metal oxides.\cite{imada}

A feature of particular interest in the organic salts is magnetic 
frustration. Since the geometric structure corresponds to an anisotropic 
triangular lattice, with inequivalent nearest neighbor hopping interactions 
$t$ and $t'$,\cite{kino,mckenzie} long-range magnetic ordering becomes 
increasingly frustrated if the lattice is nearly isotropic, giving rise 
to an exotic spin-liquid phase in the absence of symmetry breaking. 
\cite{anderson} 
Such a spin-liquid phase \cite{shimizu,kurosaki,yamashita} is realised in 
the organic insulator $\kappa$-(BEDT-TTF)$_2$Cu$_2$(CN)$_3$ (denoted below
as  $\kappa$-CN) which corresponds to $t'\approx 1.06t$, whereas 
$\kappa$-(BEDT-TTF)$_2$Cu[N(CN)$_2$]Cl (denoted as $\kappa$-Cl) 
with $t'\approx 0.75t$ is an antiferromagnetic (AF) insulator. 
\cite{lefebre,limelette}
AF order is also found in those Pd(dimt)$_2$ salts for which $0.55< t'/t < 0.85$. 
In contrast, C$_2$H$_5$(CH$_3$)$_3$P[Pd(dimt)$_2$]$_2$ with $t'=1.05t$ is a valence 
bond insulator at ambient pressure. \cite{kato}
Experiments on these kinds of two-dimensional frustrated systems have greatly 
stimulated theoretical investigations 
of the electronic and magnetic properties of anisotropic triangular lattices. 
\cite{morita,imai,onoda1,parcollet,watanabe,singh,yokoyama,kyung,%
sahebsara,koretsune,senthil,ohashi} 

The focus of the present study is the band width controlled finite temperature
phase diagram of the Hubbard model for isotropic and anisotropic triangular 
lattices. The key result is that small changes in the ratio $t'/t$ can give 
rise to fundamental changes of the phase diagram. Thus, partial and full 
magnetic frustration reveal strikingly different metal-insulator coexistence 
regions, in qualitative agreement with the experimental phase disgrams for  
$\kappa$-Cl \cite{limelette} and $\kappa$-CN. \cite{kurosaki} 

The anisotropic triangular lattice has recently been studied also by Ohashi 
{\it et al.}~\cite{ohashi} who used dynamical mean field theory\cite{dmft} (DMFT)
with a cluster extension to account for spatial fluctuations. Although at 
$t'\approx0.8t$ reentrant behavior was found as observed for $\kappa$-Cl, 
the calculated $T_c$ was much larger than the measured value. Moreover, only 
the lower boundary of the metal-insulator coexistence region was determined. 
Here, we investigate both the isotropic and anisotropic triangular lattices 
and use exact diagonalization\cite{ed} (ED) combined with cluster DMFT 
\cite{cdmft} to evaluate the upper and lower phase boundaries, $U_{c1}(T)$
and $U_{c2}(T)$, of the coexistence region. As shown below, the shape of
these boundaries, as well as the critical temperatures, are consistent with 
the experimental data for $\kappa$-Cl and $\kappa$-CN.

\section{Theory and Results}
 
The minimal model Hamiltonian that captures the interplay between geometrical 
frustration and strong Coulomb interaction present in the conducting layers 
of organic salts such as $\kappa$-Cl and $\kappa$-CN is
\begin{equation}
   H =  - \sum_{ ij\sigma} t_{ij} ( c^+_{i\sigma} c_{j\sigma} + {\rm H.c.}) 
                     + U \sum_i n_{i\uparrow} n_{i\downarrow} -
\mu\sum_{i \sigma} c^+_{i\sigma} c_{i\sigma}, 
\end{equation}
where the sum in the first term is limited to nearest neighbor sites.
The hopping integrals in a unit cell consisting of three sites are 
$t_{13}=t_{23}=t$ and $t_{12}=t'$. The band width is $W=9t$ for $t'=t$ 
and $W=8.5t$ for $t'=0.8t$. The chemical potential $\mu$ is fixed to 
give half-filling. Within cluster DMFT the interacting lattice Green's 
function in the cluster site basis is defined as
\begin{equation}
   G_{ij}(i\omega_n) = \sum_{\vec k} \left( i\omega_n + \mu - t(\vec k) - 
                   \Sigma(i\omega_n)\right)^{-1}_{ij} , \label{G}
\end{equation}
where $\vec k$ extends over the reduced Brillouin Zone and $\omega_n=(2n+1)\pi T$ 
are Matsubara frequencies. $t(\vec k)$ denotes the hopping matrix for the 
superlattice and 
$\Sigma(i\omega_n)$ represents the non-diagonal cluster self-energy matrix.
This self-energy is calculated within ED where the environment of the 
three-site cluster is replaced via a bath consisting of 6 or 9 levels, i.e.,
for a total cluster size $n_s=9$ or $n_s=12$. The calculations are carried
out in a site basis and in a mixed site/molecular orbital basis. 
Due to ED finite-size effects, these treatments give results that 
differ quantitatively. Nevertheless, the qualitative features of the 
phase diagrams are consistently reproduced by these ED versions. 
Details of the cluster ED/DMFT formalism can be found in Ref.~\cite{lie2008}.  
    
\begin{figure} 
\begin{center}
\includegraphics[width=4.2cm,height=6.5cm,angle=-90]{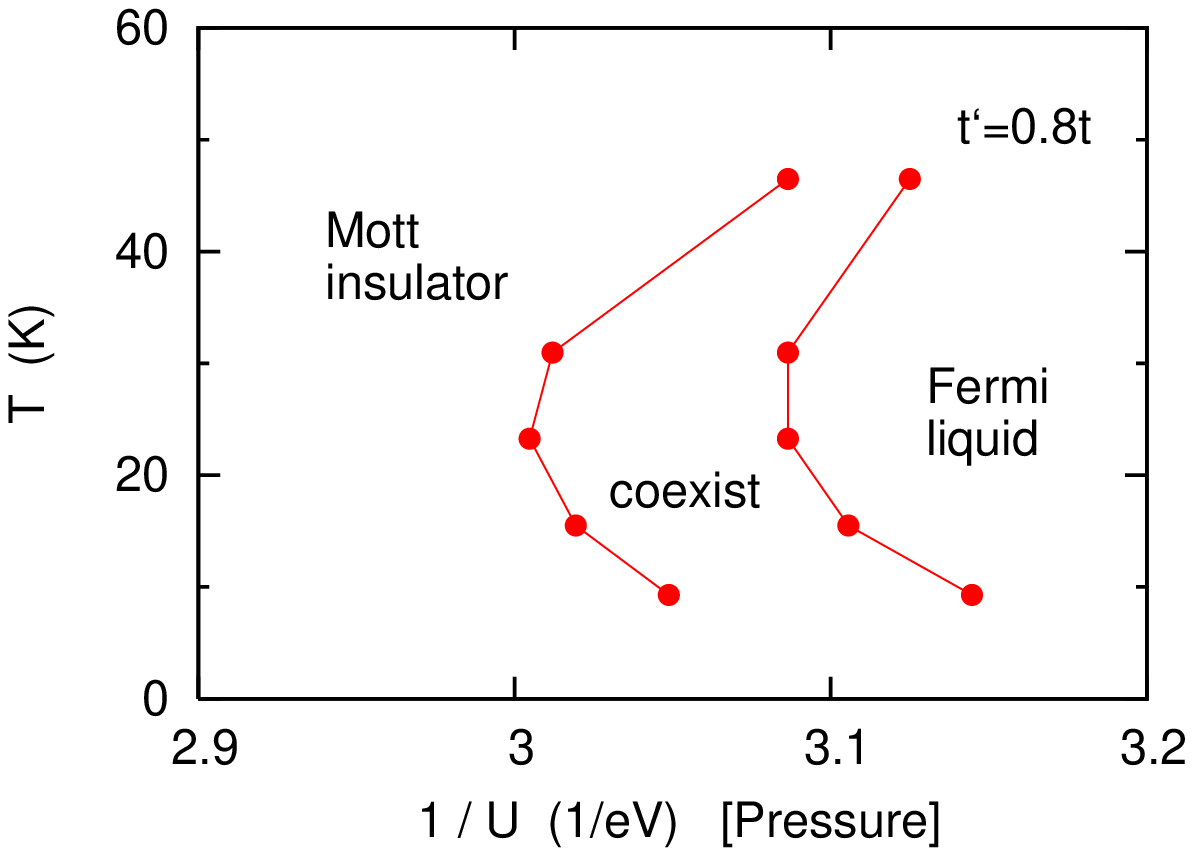}
\includegraphics[width=4.2cm,height=6.5cm,angle=-90]{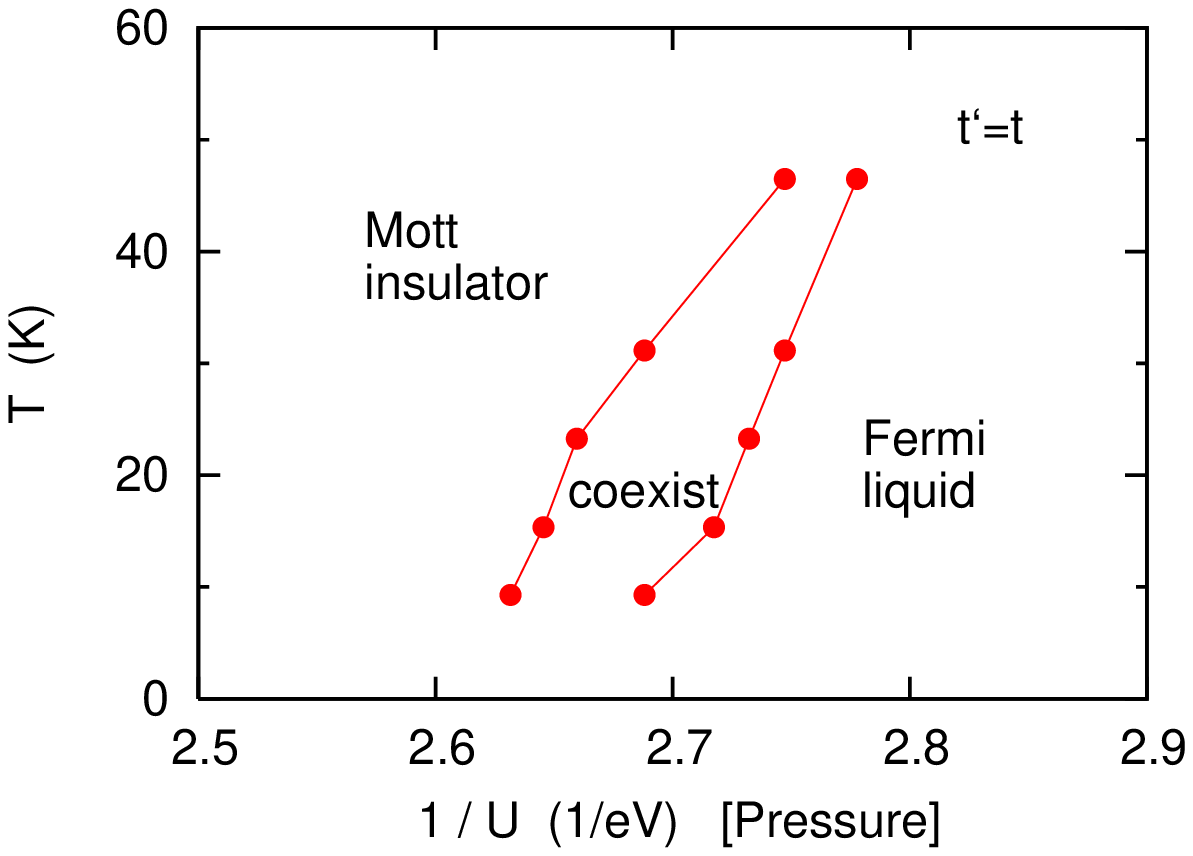}   
\end{center}\vskip-3mm
\caption{(Color online)
Phase diagrams of Hubbard model for anisotropic triangular lattice 
($t'=0.8t$) and isotropic triangular lattice ($t'=t$), evaluated
within ED cluster DMFT for $t=0.04$~eV. Plotted are the first-order 
metal-insulator phase boundaries as functions of inverse local Coulomb 
energy $U$. In the experimental setup increasing hydrostatic pressure 
$P$ implies increasing band width $W$ or decreasing $U$. The reentrant 
behavior found for $t'=0.8t$ is absent in the isotropic limit $t'=t$. 
}\label{fig1}\end{figure}

Figure~\ref{fig1}(a) shows the calculated phase diagram for the anisotropic 
lattice in the region below the critical temperature. To facilitate the 
comparison with the experimental data for $\kappa$-Cl, \cite{limelette} 
the hopping matrix elements are chosen as $t=0.04$~eV and $t'=0.8t$ to 
reproduce the single particle band width, $W=0.34$~eV. \cite{LDA} 
A similar value was used in the numerical renormalization group (NRG) 
DMFT analysis of the high-$T$ data in Ref.\cite{limelette}. 
Since the data were plotted in a $T / P$ phase diagram, we show 
the transition temperatures as functions of the inverse Coulomb energy. 
Increasing pressure $P$ implies increasing electronic band width, so that this 
measurement is equivalent to keeping $W$ fixed and reducing $U$ in the 
calculation. The phase boundaries of the coexistence region are obtained by 
carefully increasing or decreasing $U$ from the metallic or insulating domains, 
respectively.  Fig.~1(b) shows the phase diagram for the isotropic case 
corresponding to $\kappa$-CN.
 
The critical temperatures for $t'=0.8t$ and $t'=t$, $T_c\approx 50$~K~$\approx 0.11t$, 
are consistent with the measured values $T_c\approx40$~K for $\kappa$-Cl
\cite{limelette} and  $T_c\approx50$~K for $\kappa$-CN.\cite{kurosaki} 
$T_c\approx0.1t$ was recently obtained also for the fully unfrustrated square 
lattice. \cite{park} On the other hand, within Quantum Monte Carlo (QMC) DMFT 
at temperatures $T = 0.1t \ldots 1.0t$, Ohashi {\it et al.}~\cite{ohashi} 
found a much larger value, $T_c\approx 0.3t\approx140$~K.   
The experimental data and the present ED/DMFT results suggest that the metal-insulator
coexistence region is located at temperatures below those considered in 
Ref.\cite{ohashi}.  

For $t'=0.8t$, the first-order phase boundaries separating the Fermi liquid 
from the Mott insulator in Fig.~1(a) show the same kind of reentrant 
behavior as measured for $\kappa$-Cl. 
For instance, at $U=1/3$~eV and $T\approx 50$~K the system is a Mott insulator 
which turns into a Fermi liquid if $T$ is lowered to about $20$~K. Further 
reduction of $T$ reverts the system to a Mott insulator, just as seen in the 
data. (We do not consider here the antiferromagnetic insulating phase which 
is detected at even lower temperature.) Ohashi {\it et al.}~\cite{ohashi} 
found reentrant behavior at considerably higher temperatures.    

At present the origin of differences between the phase diagram for $t'=0.8t$ 
shown in Fig.~1 and the one found by Ohashi {\it et al.} is not clear. One 
reason might be that we consider a triangular lattice (3 sites per cluster) 
while in Ref.~\cite{ohashi} a square lattice with one diagonal was used
(4 sites per cluster). However, since the experimentally observed critical   
temperature is much lower than the range treated in Ref.~\cite{ohashi}, it 
would be interesting to apply continuous-time QMC to this problem in order 
to reach lower temperatures.      
     
The reentrant behavior for $t'=0.8t$ is in striking contrast to the phase 
diagram obtained for the isotropic triangular lattice shown in Fig.~1(b). 
This limit resembles more closely the phase diagram derived within 
single-site DMFT. \cite{dmft} The main effect of short-range fluctuations 
in the isotropic case is a significant lowering of the critical Coulomb energy. 
Here, $U_{c2}\approx 1/2.63$~eV~$\approx9.5t$, whereas 
$U_{c2} \approx 12t\ldots15t$ in local DMFT for the triangular lattice. 
\cite{merino06,aryanpour} Comparing Figs.~1(a) and (b), it is evident that
anisotropy causes a further lowering of the critical Coulomb energies.
This trend is consistent with $U_c\approx 6t$ for the fully unfrustrated 
square lattice \cite{park,white} which is topologically equivalent to the 
triangular lattice in the limit $t'=0$.  

It is interesting also to analyze the width of the metal-insulator coexistence 
region obtained by increasing vs. decreasing pressure. For $\kappa$-Cl, it is 
observed at $P\approx 200\ldots400$\,bar, which according to the high-$T$ NRG 
analysis corresponds to band width changes of about 2\,\%. \cite{limelette} 
The calculated coexistence region shown in Fig.~1(a) is only slightly wider 
than this experimental range.

\begin{figure}
\begin{center}
\includegraphics[width=4.2cm,height=6.5cm,angle=-90]{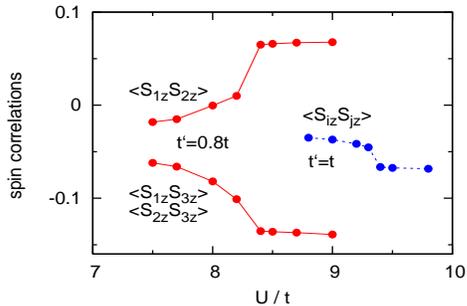}
\end{center}\vskip-5mm
\caption{(Color online) Nearest-neighbor spin correlations in 
isotropic and anisotropic triangular latticees for $T=0.05t$,
$t'=t_{12}$, $t=t_{13}=t_{23}$.
Strong enhancement of spin correlations occurs
for moderate deviations from the isotropic limit.
}\label{fig2}\end{figure}

The qualitative change in the phase diagram caused by reduced geometrical 
frustration can be understood by analyzing the magnetic properties of the 
frustrated lattice. 
In the $U\rightarrow \infty$ limit, the Hubbard model can be mapped onto 
the anisotropic Heisenberg model with $J'/J=0.64$ and $J'/J=1$ for $t'=0.8t$ 
and $t'=t$, respectively. At $T=0$, $t'=t$ yields long range antiferromagnetic
(AF) order of the 120$^0$ type, whereas $t'=0.8t$ gives rise to row-wise AF 
N\'eel order. \cite{zheng}
However, in the Heisenberg model the temperature scale for AF order in the 
isotropic triangular lattice is expected to be strongly suppressed relative 
to the square lattice.
For the Hubbard model, the cluster DMFT provides information on how the 
magnetic correlations $<S_{iz}S_{jz}>$ vary across the Mott transition 
in the isotropic case compared with $t'=0.8t$. 
The results shown in Fig.~\ref{fig2} demonstrate that spin correlations 
are strongly enhanced as the geometrical frustration is suppressed.
The isotropic lattice displays weak AF coupling for any $U$. This is in contrast 
to $t'=0.8t$, for which the weaker hopping amplitude displays ferromagnetic 
correlations whereas spins with the larger hopping amplitude are 
antiferromagnetically coupled, indicating a row-wise AF Neel arrangement of spins.
Thus, $t'=0.8t$ induces a much stronger tendency towards magnetic order than
$t'=t$, which explains why the reentrant behavior occurs for $t'=0.8t$,
 but not for $t'=t$ (see Fig.~\ref{fig1}). 
At low $T$, the electron entropy is suppressed for $t'=0.8t$ as compared to
$t'=t$. As $T$ is increased for $t'=0.8t$, the system lowers its free energy 
by transforming to a metal since the entropy of the metal exceeds that of the 
ordered insulator. At even higher temperatures the system gains entropy of 
$\log(2)$ by transforming back into a paramagnetic insulator. This result is 
analogous to the one found for the unfrustrated square lattice. \cite{park} 
In the isotropic lattice magnetic ordering is suppressed and the reentrant 
behavior disappears.
   
\begin{figure} 
\begin{center}
\includegraphics[width=4.2cm,height=6.5cm,angle=-90]{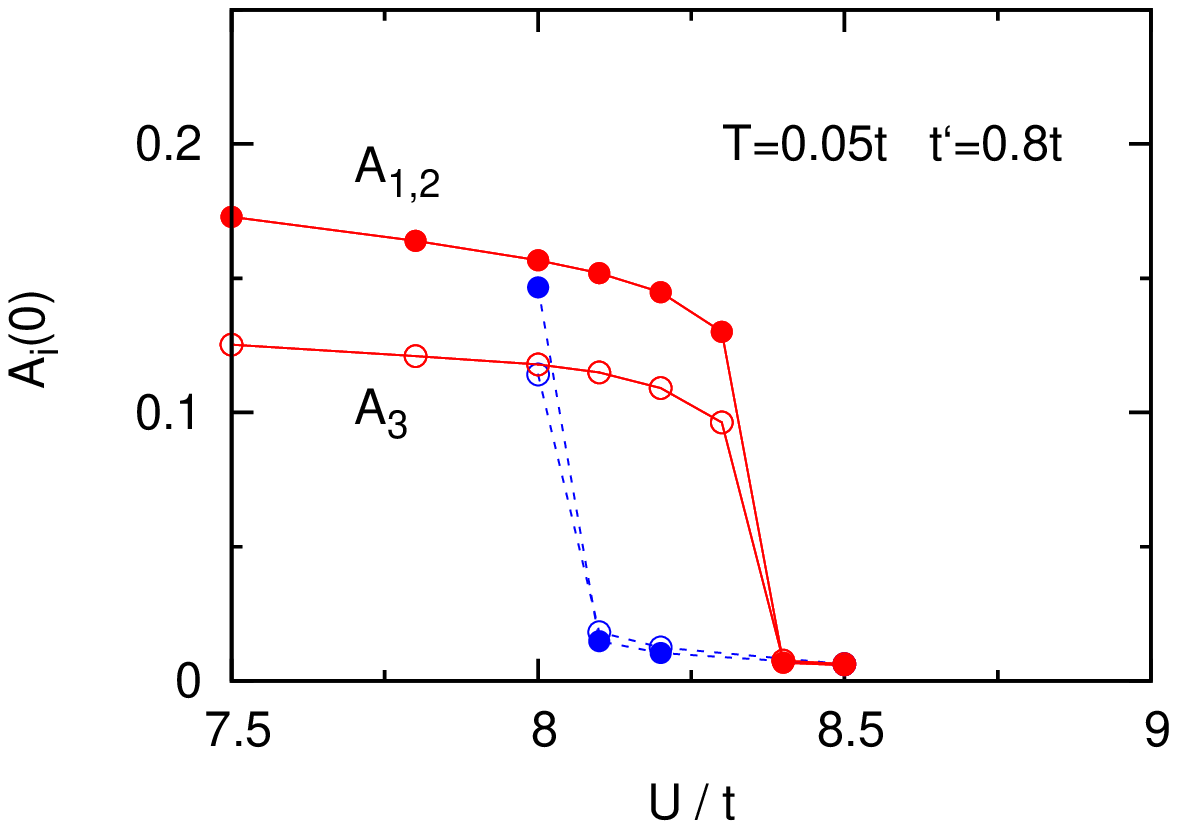}
\includegraphics[width=4.2cm,height=6.5cm,angle=-90]{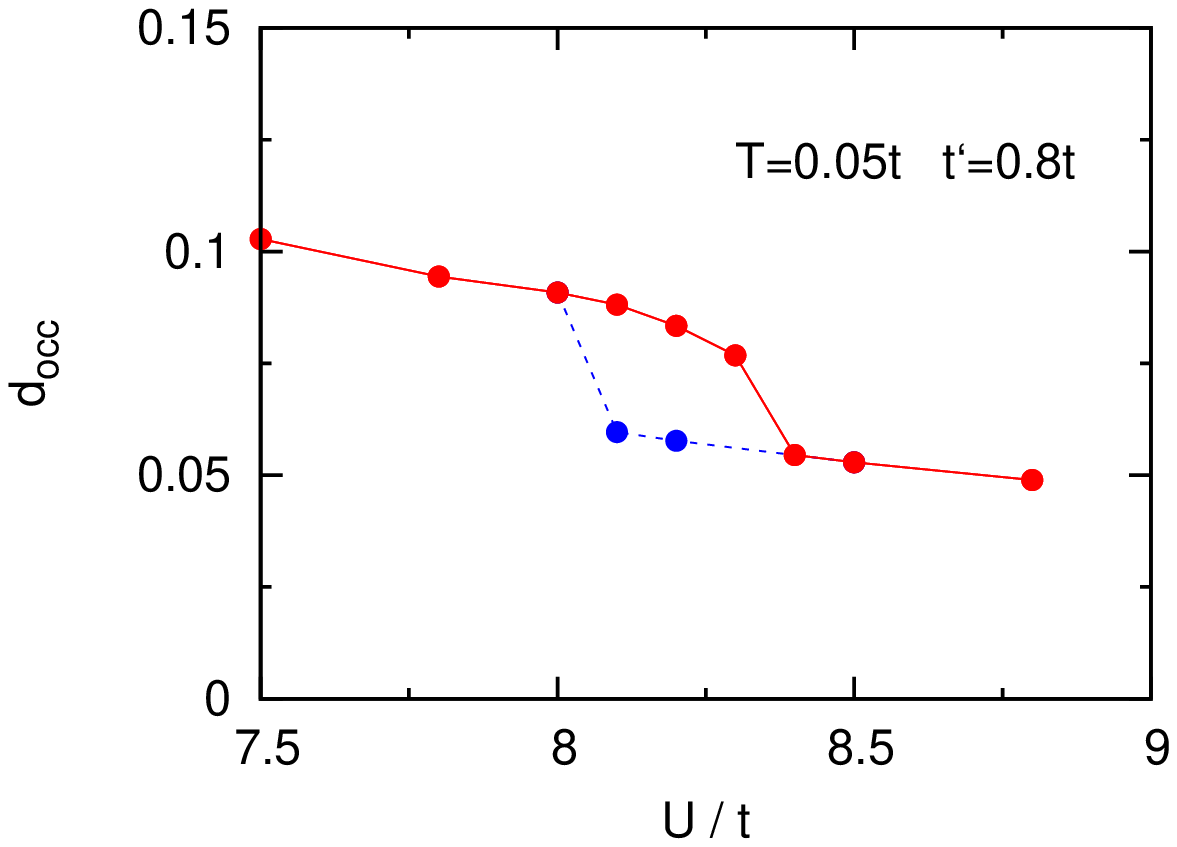}
\end{center}\vskip-5mm
\caption{(Color online)
Hysteresis behavior of spectral weights $A_i(0)$ of cluster sites 
at $E_F=0$ and average double occupancy $d_{occ}$ as functions of 
Coulomb energy for anisotropic triangular lattice 
($t=1$, $t'=0.8t$,  $T=0.05t$). 
Red (blue) curves: increasing (decreasing) $U$.
}\label{fig3}\end{figure}

\begin{figure}[b!] 
\begin{center}
\includegraphics[width=4.2cm,height=6.5cm,angle=-90]{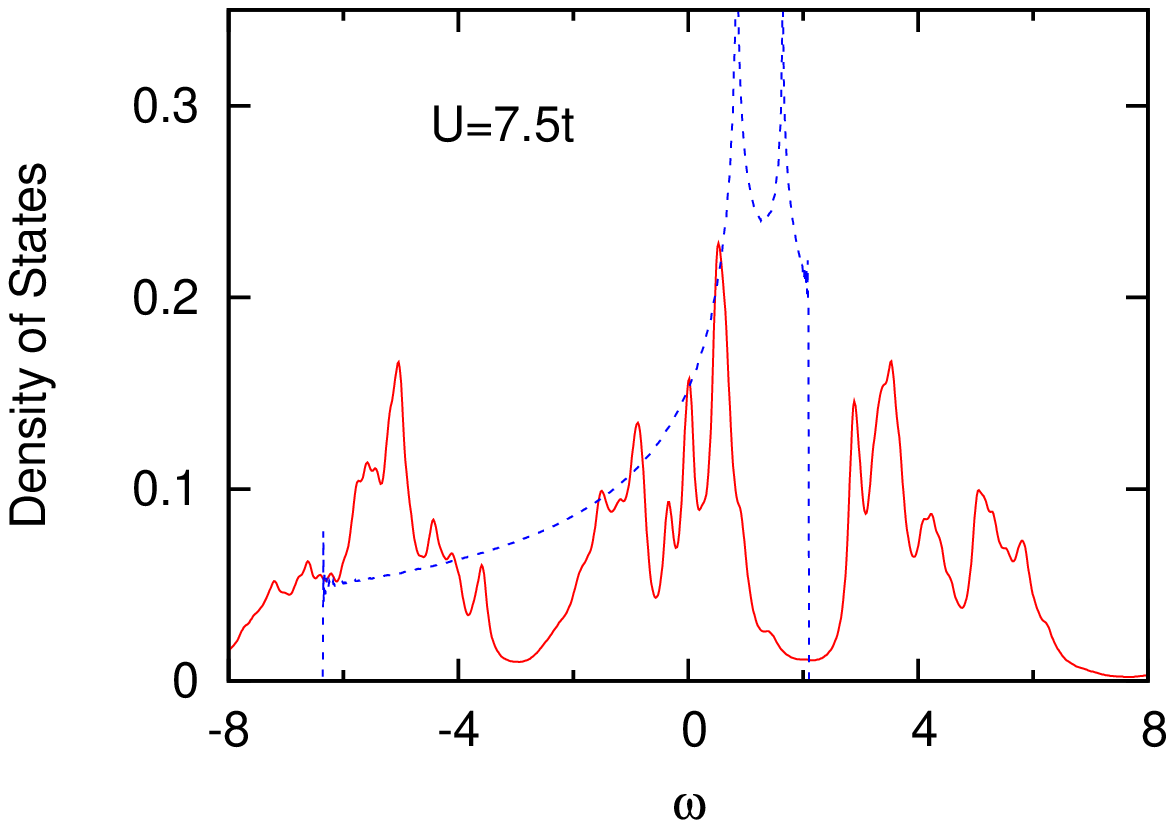}
\includegraphics[width=4.2cm,height=6.5cm,angle=-90]{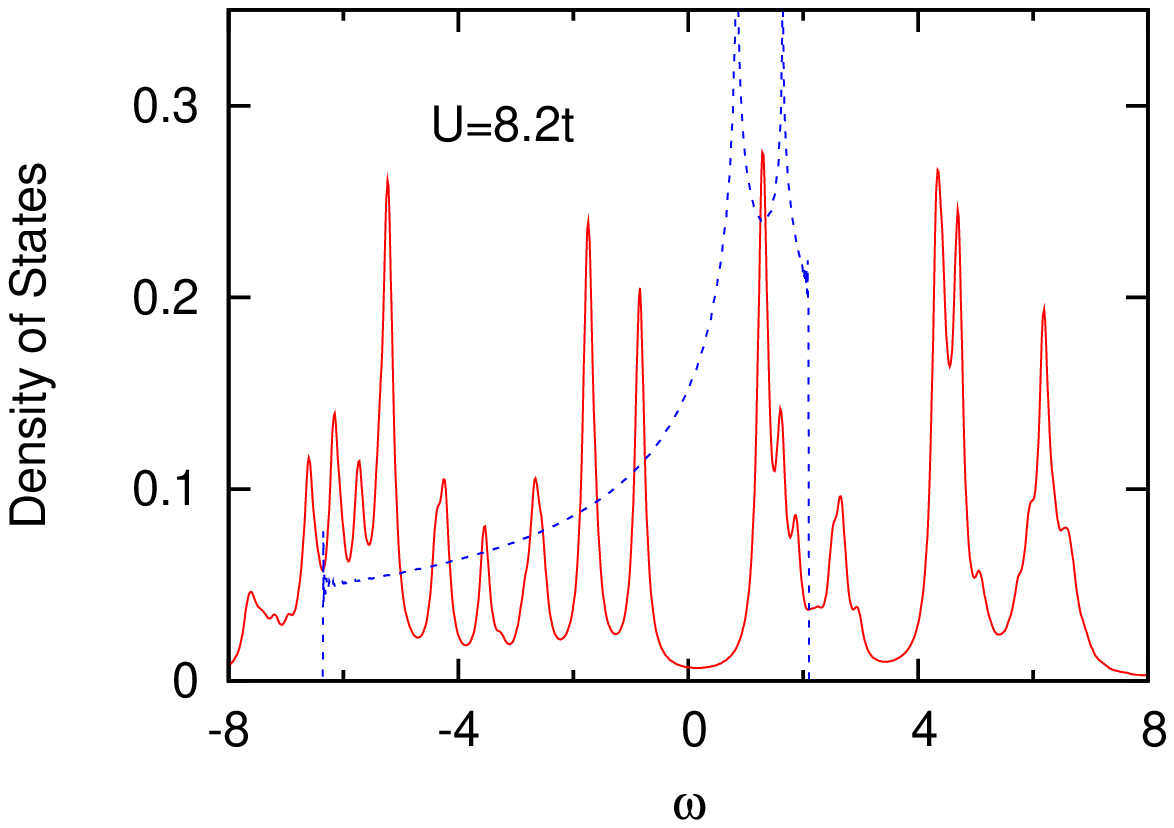}
\end{center}\vskip-5mm
\caption{(Color online)
Average spectral distributions of cluster sites below and above 
the Mott transition for the anisotropic triangular lattice 
($t=1$, $t'=0.8t$,  $T=0.02t$) for Coulomb energies 
$U=7.5t$ and $U=8.2t$.
The bare density of states is shown by the dashed blue curve.
}\label{fig4}\end{figure}

To illustrate the first-order nature of the metal-insulator transition 
we show in the upper panel of Fig.~\ref{fig3} the spectral weights of the 
cluster sites at $E_F=0$ as functions of $U$.
The lower panel shows the average double occupancy 
$d_{\rm occ}=\sum_i\langle n_{i\uparrow}n_{i\downarrow}\rangle/3$. 
Both quantities exhibit hysteresis for increasing and decreasing $U$, 
indicating coexistence of metallic and insulating solutions. 

Finally, Fig.~\ref{fig4} shows the spectral densities at Coulomb energies below 
and above the Mott transition for $T=0.02t$ and $t'=0.8t$. Plotted is the average
over the three inequivalent sites within the unit cell. Since we are here 
concerned with the metal-insulator transition we give the ED cluster spectra
which can be evaluated without requiring extrapolation from Matsubara to real
frequencies. In the metallic phase the spectra reveal large quasi-particle
weight at low frequencies as well as upper and lower Hubbard bands at high
frequencies. The insulating phase exhibits a Mott gap, as well as
pronounced spectral weight in the region of the Hubbard bands. Qualitatively
similar features are also seen for the unfrustrated square lattice. 
\cite{park,zhang}

\section{Conclusion}

In conclusion, the phase diagrams of the Hubbard model for the isotropic and
anisotropic triangular lattices have been determined within cluster DMFT and 
exact diagonalization. For moderate frustration, $t'=0.8t$, reentrant behavior 
is found and the phase boundaries of the metal-insulator coexistence region are 
in qualitative agreement with the $T / P$ phase diagram observed experimentally
for the anisotropic organic salt $\kappa$-Cl. The reentrant behavior disappears 
in the fully frustrated limit, $t'=t$, in agreement with measurements on the 
nearly isotropic compound $\kappa$-CN. The phase diagram then bears overall 
resemblance to the one obtained within local DMFT, i.e., in the absence of 
inter-site correlations. The critical temperatures, $T_c\approx 50$~K for
the isotropic and anisotropic lattices, are consistent with the data for 
$\kappa$-CN and $\kappa$-Cl. These results should also be relevant for the 
phase diagram of [Pd(dimt)$_2$]$_2$ salts exhibiting small deviations from 
the isotropic lattice. \cite{kato2} 

Computational work (A.L.) was carried out on the Juelich JUMP. 
J.M. thanks MCI for financial support: CTQ2008-06720-C02-02/BQU.

\end{document}